\documentclass[manuscript=fullresearchpaper]{beilstein}
\usepackage{amsfonts}
\usepackage{amssymb}
\usepackage{bm}
\usepackage{amsmath}
\usepackage{graphicx}
\usepackage{tabularx}
\usepackage{dcolumn}
\usepackage{xcolor}
\begin{document}
\nolinenumbers
\title{Exploiting the Hierarchical Morphology of Single-Walled and Multi-Walled Carbon Nanotube Films for Highly Hydrophobic Coatings}
\author*{Francesco De Nicola}{fdenicola@roma2.infn.it}
\affiliation{Dipartimento di Fisica, Universit\'a di Roma Tor Vergata, Via della Ricerca Scientifica 1, 00133 Roma, Italy}
\affiliation{Istituto Nazionale di Fisica Nucleare, Universit\'a di Roma Tor Vergata (INFN-Roma Tor Vergata), Via della Ricerca Scientifica 1, 00133 Roma, Italy}
\author[1,2]{Paola Castrucci}
\author[1,2]{Manuela Scarselli}
\author{Francesca Nanni}
\affiliation{Dipartimento di Ingegneria dell'Impresa, Universit\'a di Roma Tor Vergata (INSTM-UdR Roma Tor Vergata), Via del Politecnico 1, 00133 Roma, Italy}
\author{Ilaria Cacciotti}
\affiliation{Universit\'a di Roma Niccol\`o Cusano (INSTM-UdR), Via Don Carlo Gnocchi 3, 00166 Roma, Italy}
\author[1,2]{Maurizio De Crescenzi}
\affiliation{Istituto di Struttura della Materia, Consiglio Nazionale delle Ricerche (ISM-CNR), Via del Fosso del Cavaliere 100, 00100 Roma, Italy}
\maketitle
\begin{abstract}
Self-assembly hierarchical solid surfaces are very interesting for wetting phenomena, as observed in a variety of natural and artificial surfaces. Here, we report single-walled (SWCNT) and multi-walled carbon nanotube (MWCNT) thin films realized by a simple, rapid, reproducible, and inexpensive filtration process from an aqueous dispersion, then deposited by dry-transfer printing method on glass, at room temperature. Furthermore, the investigation of carbon nanotube films by scanning electron microscopy (SEM) reveals the multi-scale hierarchical morphology of self-assembly carbon nanotube random networks. Moreover, contact angle measurements show that hierarchical SWCNT/MWCNT composite surfaces exhibit a higher hydrophobic behavior (up to 137$^{\circ}$) than bare SWCNT (110$^{\circ}$) and MWCNT (97$^{\circ}$) coatings, thereby confirming the enhancement produced by the surface hierarchical morphology.
\end{abstract}
\section{Introduction}
\label{sec:intro}
\indent In general, surface morphology \cite{Wenzel1936} is a crucial parameter for the fabrication of artificial hydrophobic surfaces and may be enhanced especially by hierarchical \cite{Feng2008,Sun2005,XingJiuHuang2007,Jung2009,Egatz-Gomez2012,Bittoun2012} and fractal structures \cite{Shibuichi1996,Bittoun2012}, possibly allowing air pocket formation to further repel water penetration \cite{Cassie1944}.\\
\indent In particular, surface hierarchical morphology is a recent concept introduced to explain the wetting properties of surfaces such as plant leaves \cite{Sun2005,Feng2008}, bird feathers \cite{Bormashenko2007}, and insect legs \cite{Gao2004}. These surfaces are made of a hierarchical micro- and nano-morphology which improves their wettability.\\ 
\indent It is indeed well-established \cite{Lafuma2003,Giacomello2012}, that in composite rough surfaces hierarchical morphology may induce a wetting transition from Wenzel \cite{Wenzel1936} to Cassie-Baxter \cite{Cassie1944} state, owing to air trapping. Moreover, this transition may occur by passing through thermodynamically metastable states \cite{Giacomello2012,Giacomello2012a,Savoy2012,Murakami2014}, where the free energy surface presents one absolute minimum and one or more local minima separated from the former by large free energy barriers, as compared to the thermal energy. Metastability can also have a technological importance, as in practice, it represents a way of extending the range of stability of the Cassie-Baxter state \cite{Giacomello2012a,Bico2002}. Conversely, a negative consequence of metastability is that it might prevent or slow down the transition between Wenzel and Cassie-Baxter states \cite{Giacomello2012a,Bico2002}.\\
\indent Moreover, biomimetics \cite{Bhush2009,Bar-Cohen2005} may be exploited in order to realize cutting edge artificial surfaces \cite{Feng2008,Sun2005,Jung2009} mimicking natural ones; making in this way these surfaces ideal for hydrophobic (lipophilic) and/or hydrophilic (lipophobic) applications.\\
\indent With the same spirit, here we report the fabrication of highly hydrophobic coatings by self-assembling SWCNTs on MWCNTs. Since the former have a smaller characteristic dimension than the latter (about one order of magnitude), we observed that a surface hierarchy naturally occurs by depositing layer by layer a SWCNT film upon a MWCNT film. The particular surface two-fold hierarchical morphology, resembling that observed in lotus leaves \cite{Sun2005} and rose petals \cite{Feng2008} where micro-papillae are made of nano-papillae, improves the hydrophobic behavior of carbon nanotube coatings compared to bare SWCNT and MWCNT films. Moreover, we report for the first time the experimental Wenzel-Cassie-Baxter phase diagram \cite{Bico2002,Lafuma2003,Shibuichi1996} for a carbon nanotube surface, showing that the transition between the Wenzel and Cassie-Baxter states occurs by passing through metastable states.\\
\indent Generally, carbon nanotubes \cite{Iijima1991,Jorio2008} are the one-dimensional allotropic form of carbon with cylindrical symmetry and a $sp^{2}$ lattice. Carbon nanotubes may be single-walled or multi-walled depending on the number of coaxially arranged graphite planes. Moreover, owing to their honeycomb lattice, carbon nanotubes are inherently hydrophilic (graphite contact angle with water $\approx86^{\circ}$ \cite{Adamson1997}) but apolar. However, by surface functionalization or textured arrangement it can be possible realizing carbon nanotube films which offer versatility, high stability, and multi-functionality owing to their exceptionally unique properties \cite{Jorio2008}, making their usage widespread in hydrophobic surface realizations \cite{Georgakilas2008,XingJiuHuang2007,Wang2008,Yang2010,Kakade2008,Nasibulin2011,Wang2010,Wang2011,Li2010a,Jung2009,Wang2007,Zhang2009,Bu2010,Kakade2008a,Li2002}.\\
\indent Furthermore, self-assembly hierarchical nano-structured materials \cite{Li2006,Chakrapani2004,Bohn2005,Fusi2011}, are nowadays investigated as a consequence of their tunable peculiar properties, easy, high reproducible, and low-cost fabrication. In addition, they are ideal low-dimensional materials for the fabrication of high aspect ratio and large area devices \cite{Sauvage2010}.
\section{Results and Discussion}
\label{sec:results}
\begin{figure*}
	\centering
		\includegraphics[width=16.4cm,keepaspectratio]{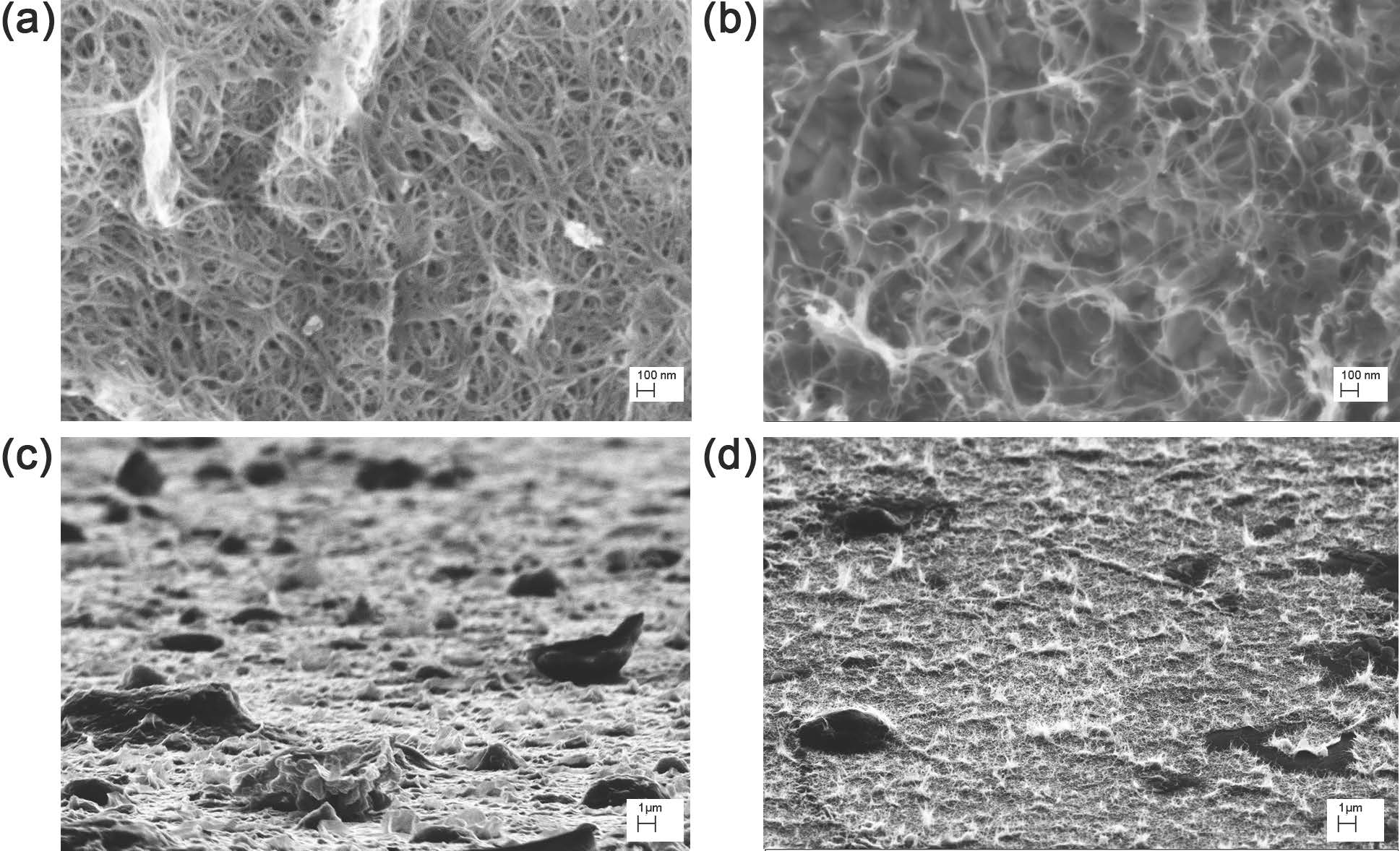}
	\caption{Scanning electron micrographs of SWCNT (a,c) and MWCNT (b,d) films at different magnifications
200,000$\times$ (a,b), and 10,000$\times$ (c,d). In the images taken at grazing incidence (c,d) it is possible to observe that SWCNTs (c) self-assembly in ripples forming several micro-structures, while MWCNTs (d) just aligned in the out-of-plane vertical direction. (d) Black areas are holes in the film.}
	\label{fig:Figure1}
\end{figure*}

\begin{figure*}
	\centering
		\includegraphics[width=16.4cm,keepaspectratio]{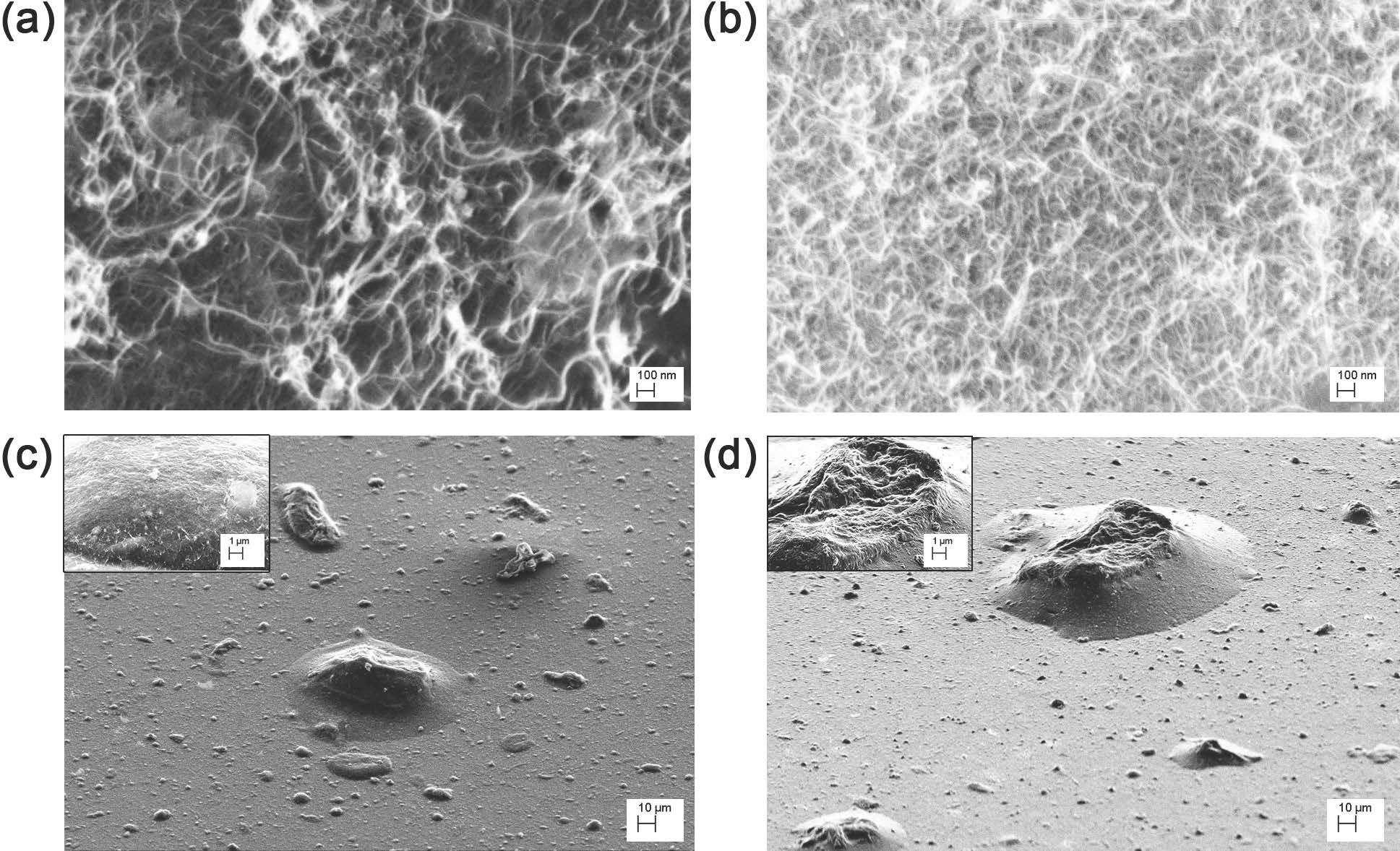}
	\caption{Scanning electron micrographs of SWCNT/MWCNT (a,c) and MWCNT/SWCNT (b,d) films at different magnifications 200,000$\times$ (a,b), 1,000$\times$ (c,d), and 10,000$\times$ (c,d insets). In the images taken at grazing incidence (c,d and insets), it is possible to observe that in both cases the self-assembly forms several huge micro-structures. (c,d insets) Details of the micro-structures showing a hierarchical morphology very similar to that of lotus leaves and rose petals.}
	\label{fig:Figure2}
\end{figure*}

\begin{figure}
	\centering
		\includegraphics[width=8.2cm,keepaspectratio]{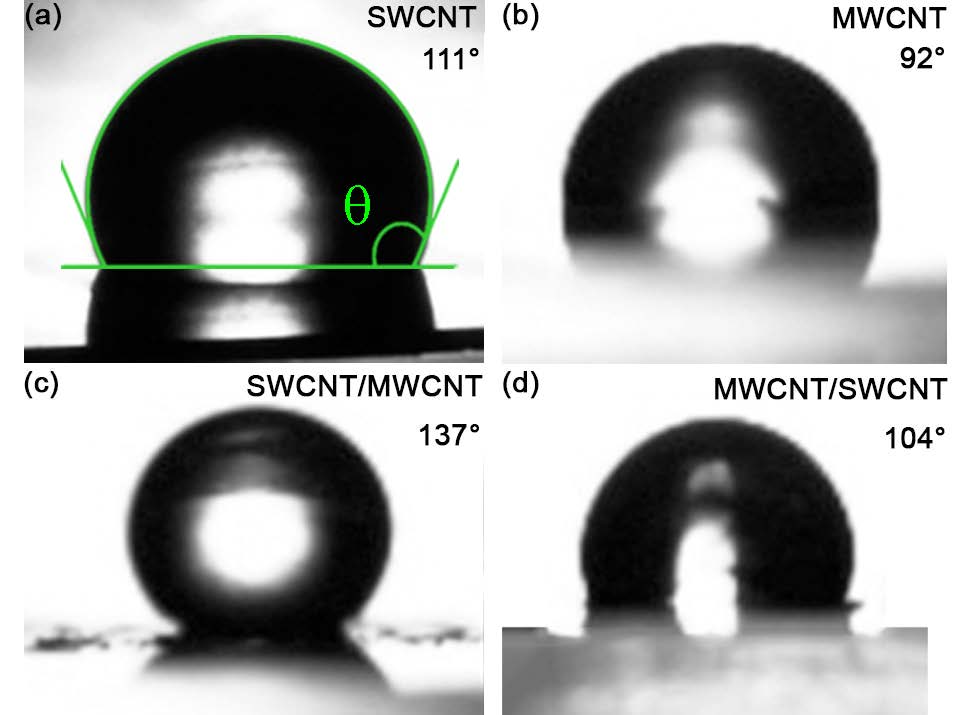}
	\caption{Water droplets cast on SWCNT (a), MWCNT (b), SWCNT/MWCNT (c), and MWCNT/SWCNT (d) films. Owing to the rough and porous surface of the samples, water drops exhibit different contact angle values, depending on the surface point they are cast. In this case, the contact angle can be only defined on average.}
	\label{fig:Figure3}
\end{figure}

\begin{figure} 
	\centering
		\includegraphics[width=16.4cm,keepaspectratio]{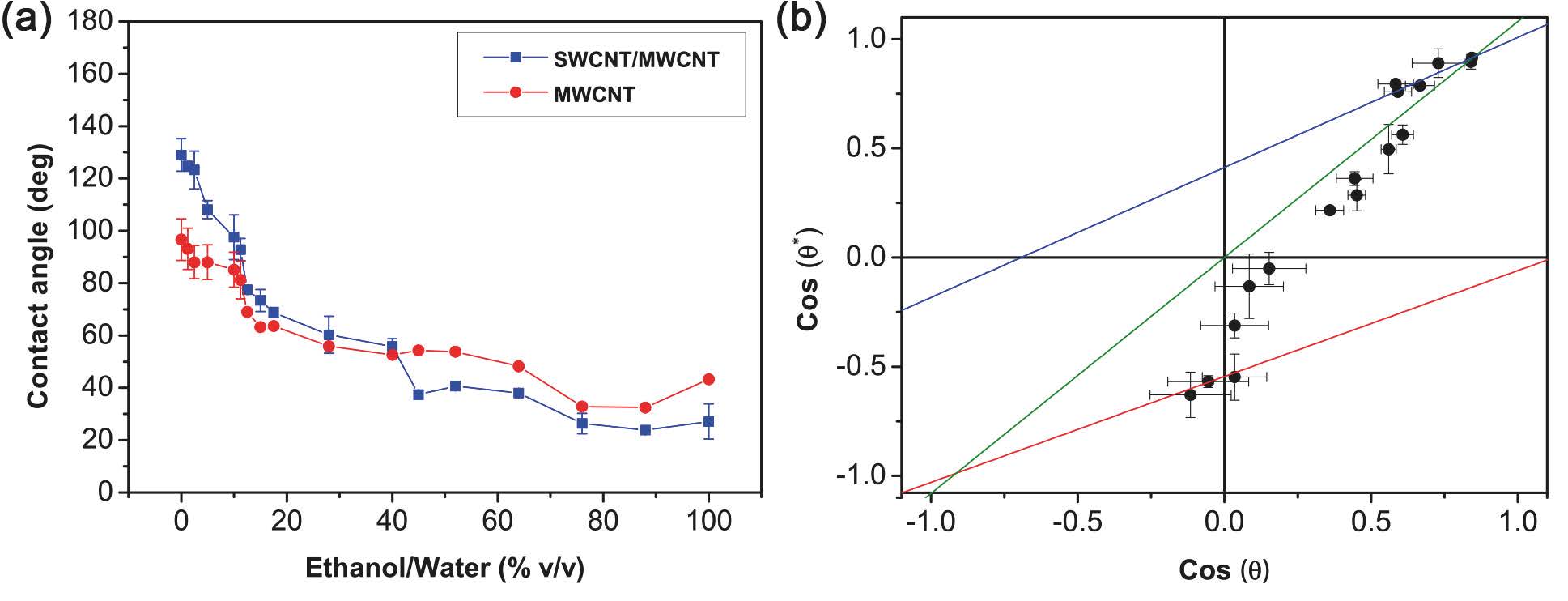}
	\caption{(a) Contact angle of the SWCNT/MWCNT (blue squares) and MWCNT (red dots) films as a function of ethanol concentration in water. (b) Wenzel-Cassie-Baxter phase diagram of the SWCNT/MWCNT surface respect to the MWCNT surface. Wetting states are studied changing the liquid surface tension by adding different ethanol concentrations in water. Wenzel regime (green solid line) fit reports a roughness factor $r=1.08\pm0.01$, while lipophilic (blue solid line) and hydrophobic (red solid line) Cassie-Baxter regime fits report respectively a liquid fraction $\phi_{+}=0.41\pm0.04$ and an air fraction $\phi_{-}=0.54\pm0.02$. The Wenzel-Cassie-Baxter transition point in the hydrophobic regime is the intersection between the red and green solid lines, while in the lipophilic regime it is the intersection between the blue and green solid lines. Error bars are standard deviations.}
	\label{fig:Figure4}
\end{figure}

\begin{figure} 
	\centering
		\includegraphics[width=8.2cm,keepaspectratio]{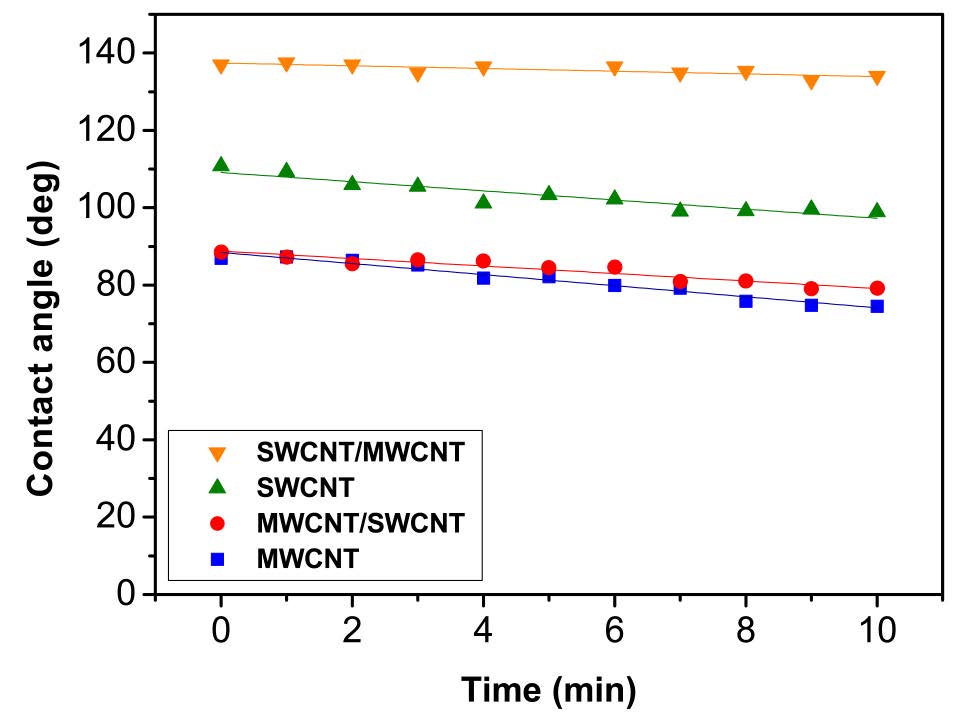}
	\caption{Variations of the contact angle as a function of the elapsed time from drop cast on the porous SWCNT (green triangles), MWCNT (blue squares), SWCNT/MWCNT (orange inverted triangles), MWCNT/SWCNT (red dots) films. The quite constant trend of the SWCNT/MWCNT contact angle value shows the stability in time of the carbon nanotube hydrophobic Cassie-Baxter metastable state.}
	\label{fig:Figure5}
\end{figure}

\indent The films obtained from the process described in the Experimental section are porous random networks of SWCNTs and MWCNTs exhibiting a hierarchical morphology made of micro- and nano-structures, as evident from SEM micrographs in Figure \ref{fig:Figure1}a-d. From SEM image analysis (see Experimental section), we estimated the pore radius $\rho$ and the bundle diameter $d$ of the SWCNT and MWCNT random networks. The obtained results are reported in Table \ref{tab:table1} together with the SWCNT micro-structure area $S$ and height $h$. However, in the case of MWCNT films, no micro-structures were observed. It is worth noting that the characteristic dimension $d$ of MWCNTs is bigger about one order than that of SWCNTs.\\
\indent In particular, we considered the micro-structures shown in Figure \ref{fig:Figure1}c as ripples randomly distributed within the film. Such self-assembly occurs by an out-of-plane bending process during evaporative drying of single-walled carbon nanotube film during its preparation \cite{Li2006,Chakrapani2004,DeNicola2014}. The dry-induced, out-of-plane assembly is the result of the competition between attractive capillary forces and bending stress due to the elasticity of SWCNT film. Once the liquid is completely evaporated, a pattern of micrometer-sized randomly shaped islands is formed. If after complete evaporation there is a balance between adhesion and elastic energy, the micro-structures are in a stable bended configuration, with respect to further wetting-dewetting cycles. This self-assembly leads to an intrinsic hierarchical micro-structured (ripples) and nano-structured (carbon nanotubes) roughness able to enhance SWCNT film wetting properties. Conversely, the MWCNT sample (Figure \ref{fig:Figure1}d)
\\
just aligned in the out-of-plane vertical direction.\\
\indent Furthermore, we induced an extrinsic hierarchical architecture by depositing a SWCNT film on a MWCNT film (SWCNT/MWCNT) and in reverse order (MWCNT/SWCNT), as shown in Figure \ref{fig:Figure2}a-d. From SEM image analysis, we obtained the two film pore diameters, micro-structure areas and heights, as reported in Table \ref{tab:table1}. In both the cases, a self-assembly led to the formation of several huge micro-structures, as compared to the those of the SWCNT films.\\
\indent Moreover, in Figure \ref{fig:Figure3}a,b images of water droplets cast on our SWCNT and MWCNT films are reported, with average contact angle values $\theta=110^{\circ}\pm3^{\circ}$ and $\theta=97^{\circ}\pm8^{\circ}$, respectively. These results can
be ascribed to the particular morphology of both the films induced by the inherent properties of the carbon nanotubes (e.g., self-assembly, nanotube diameter and spatial orientation) and film preparation method. We also found that for the SWCNT/MWCNT sample the extrinsic surface hierarchy improved the MWCNT sample hydrophobicity, exhibiting a highly hydrophobic average contact angle value $\theta=129^{\circ}\pm8^{\circ}$ (Figure \ref{fig:Figure3}c), comparable to $\approx108^{\circ}$-$118^{\circ}$\cite{Adamson1997,Clark2012} of PTFE (Teflon). Conversely, for the MWCNT/SWCNT sample (Figure \ref{fig:Figure3}d) a slightly decrease of the average contact angle value ($\theta=103^{\circ}\pm7^{\circ}$) with respect to the bare SWCNT sample was encountered. 
Our results, summarized in Table \ref{tab:table1}, may be interpreted on the basis of the micro-structure characteristic dimensions $S$ and $h$.
\begin{table}
\caption{Experimental results of SEM analysis and contact angle measurements.}
\label{tab:table1}
\centering
\begin{tabular}{|c|c|c|c|c|c|}\hline
\textbf{Sample} & $\bm{\rho}$ \textbf{(nm)} & $\bm{d}$ \textbf{(nm)} & $\bm{S}$ ($\bm{\mu}$\textbf{m}$^{\bm{2}}$) & $\bm{h}$ ($\bm{\mu}$\textbf{m)} & $\bm{\theta}$ \textbf{(deg)}\\\hline
SWCNT & 2-8 & 4-8 & 0.003-0.007 & 1.6-11.7 & $110\pm3$\\\hline
MWCNT & 40-44 & 34-84 & - & - & $97\pm8$\\\hline
SWCNT/MWCNT & 47-51 & - & 7.3-13.7 & 3.2-61.6 & $129\pm8$\\\hline
MWCNT/SWCNT & 5-7 & - & 7.7-14.3 & 2.3-76.2 & $103\pm7$\\\hline
\end{tabular}
\end{table}
In both the SWCNT/MWCNT and MWCNT/SWCNT samples the micro-structure characteristic dimensions are comparable with those of lotus and rose micro-papillae \cite{Feng2008,Sun2005}. Nevertheless, in the latter the extrinsic hierarchical morphology is reversed (bigger MWCNT scale superimposed on the smaller SWCNT scale), thereby losing the hierarchical fakir effect \cite{Quere2002,Bittoun2012}. Therefore, the SWCNT/MWCNT sample has the best hydrophobic behavior because is the most biomimetic. We remark that the large deviation from the contact angle average value is due to the highly rough and porous surface of our samples. In addition no roll-off angle value could be measured for each film, evidently due to the high contact angle hysteresis, which pinned the droplets to the surface \cite{Feng2008}.\\
\indent In order to better understand the origin of the enhancement provided by the SWCNT/MWCNT film over the MWCNT film, we characterized the wetting state of the former composite surface with respect to the latter. In Figure \ref{fig:Figure4}a, we report the contact angle of both the films as a function of the concentration in volume percent of ethanol in water. It is possible to observe that since ethanol has a lower liquid-vapor surface tension ($\gamma_{LV}=22$ mJ m$^{-2}$) than water ($\gamma_{LV}=72$ mJ m$^{-2}$), the higher the ethanol concentration in water, the lower the solution surface tension. Furthermore, the contact angle is generally proportional to the liquid surface tension by the Young's relation
\begin{equation}
	\cos{\theta}=\frac{\gamma_{SV}-\gamma_{SL}}{\gamma_{LV}},
\end{equation}
where $\gamma_{SV}$ and $\gamma_{SL}$ are the solid-vapor and solid-liquid surface tensions. Therefore, also the carbon nanotube film contact angles decrease with the decrease in liquid droplet surface tension. This phenomenon is connected to the lipophilicity of the carbon nanotube apolar surface. Indeed, on our carbon nanotube films no contact angle ($\theta\approx0$) can be measured for pure ethanol droplets. Therefore, we could investigate all the wetting phenomena occurring on our carbon nanotube surface, exploring all the wetting states. We further noted that for $\theta\approx56^{\circ}$ ($\cos{\theta}\approx0.56$) there is an intersection point between the two curves in Figure \ref{fig:Figure4}a, beyond that the SWCNT/MWCNT surface becomes more lipophilic than the MWCNT surface. That point corresponds to the Wenzel to Cassie-Baxter transition point in the lipophilic region of the Wenzel-Cassie-Baxter phase diagram, as confirmed from the plot (first quadrant) in Figure \ref{fig:Figure4}b. However, the plot in Figure \ref{fig:Figure4}b shows that the transition occurs by passing through metastable states with an abrupt change in the wetting state. We fitted our data with the lipophilic Cassie-Baxter's equation \cite{Cassie1944}
\begin{equation}
\cos{\theta^{\ast}}=\left(1-\phi_{+}\right)\cos{\theta}+\phi_{+},\qquad
1=\phi+\phi_{+},
\label{eq:cassie}
\end{equation}
with $\phi$ the surface solid fraction, $\phi_{+}$ the surface fraction wet by liquid, $\theta^{\ast}$ the SWCNT/MWCNT surface contact angle and $\theta$ the MWCNT surface contact angle. We obtained from fit a liquid fraction $\phi_{+}=0.41\pm0.04$ in contact with the droplet. However, we remark that these metastable Cassie-Baxter states coexist with the Wenzel states, which are stable because lower in surface free energy.\\
\indent Moreover we fitted our data in Figure \ref{fig:Figure4}b with Wenzel's equation \cite{Wenzel1936}
\begin{equation}
\cos{\theta^{\ast}}=r\cos{\theta}, \qquad r\geq1,
\end{equation}
where $r$ is the roughness factor (i.e, the ratio between the actual wet surface area and its projection on the plane). Interestingly, the fit returned $r=1.08\pm0.01$, which means that substantially the SWCNT/MWCNT sample has the same roughness of the MWCNT sample. It is worth noting that in our case $r\approx1$ does not mean that the surface is smooth, because we are not comparing the SWCNT/MWCNT with its corresponding smooth surface with the same chemistry, such as plain graphite. However, in the latter case we would have had a high roughness factor \cite{DeNicola2014}. Therefore, we can exclude a roughness enhancement, which we did not observe, as the reason of a such improvement in the SWCNT/MWCNT sample hydrophobic behavior over the MWCNT sample. In addition, by the relation \cite{Gennes2003}
\begin{equation}
	\cos{\theta^{\prime}}=\frac{1-\phi_{+}}{r-\phi_{+}},
\end{equation}
we can infer that the lipophilic Wenzel-Cassie-Baxter transition point is $\cos{\theta^{\prime}}=0.88$ (the intersection between the blue and green solid lines in Figure \ref{fig:Figure4}b), which is beyond the measured data, thus confirming that the achieved lipophilic Cassie-Baxter states are metastable.\\ 
\indent Conversely, in the hydrophobic region (third quadrant of the plot) we observe a sharp discontinuity beyond $\cos{\theta}=0$, confirming again that the transition between the Wenzel and Cassie-Baxter states is not continuous, but it undergoes metastable states which slow down the dewetting process. Actually, by fitting our data in Figure \ref{fig:Figure4}b with the hydrophobic Cassie-Baxter's equation
\begin{equation}
	\cos{\theta^{\ast}}=\left(1-\phi_{-}\right)\cos{\theta}-\phi_{-},\qquad
	1=\phi+\phi_{-},
\end{equation}
we obtained an air surface fraction $\phi_{-}=0.54\pm0.02$ below the liquid droplet. Furthermore, by the relation \cite{Gennes2003}
\begin{equation}
	\cos{\theta^{\prime\prime}}=\frac{\phi_{-}-1}{r-\phi_{-}},
\end{equation}
we can infer that the hydrophobic Wenzel-Cassie-Baxter transition point is $\cos{\theta^{\prime\prime}}=-0.85$ (the intersection between the red and green solid lines in Figure \ref{fig:Figure4}b), which is beyond the measured data, thus confirming that the achieved hydrophobic Cassie-Baxter states are metastable. Nevertheless, this result suggests a consistent air pocket formation \cite{Cassie1944}. 
Therefore, we can assert that the only cause of the SWCNT/MWCNT film improved hydrophobicity/lipophilicity over the MWCNT film, is the fakir effect induced by the two-fold hierarchical morphology given by the SWCNT film superimposed on the MWCNT film. This particular morphology induces air pocket formation when the interaction with the liquid is hydrophobic, otherwise it favourites the formation of a precursor liquid film \cite{Gennes2003} that enhances the wettability of the carbon nanotube surface, when the interaction with the liquid is lipophilic.\\
\indent Furthermore, we studied the stability in time of our carbon nanotube films by performing suction experiments. Figure \ref{fig:Figure5} reports the variations of contact angle value as a function of the elapsed time from drop cast on the SWCNT, MWCNT, SWCNT/MWCNT, and MWCNT/SWCNT coatings. In such suction experiment, we show that although samples are porous, the contact angle trend is quite constant. In particular, we demonstrated the stability in time of the hydrophobic Cassie-Baxter metastable state for the SWCNT/MWCNT sample. However, the linear slightly decrease of the contact angle in time is both due to liquid evaporation and suction by the porous films. Our results are particularly remarkable, since the water contact angle of carbon nanotube films has been reported \cite{Huang2005} so far to linearly decrease with time, from an initial value of $\approx$ 146$^{\circ}$ to $\approx$ 0 within 15 min.
\section{Conclusion}
\label{sec:conclusion}
\indent Single-walled and multi-walled carbon nanotube films were prepared by vacuum filtration of an aqueous dispersion. Such coatings were deposited by dry-transfer printing on glass, at room temperature. Furthermore, SEM images revealed the intrinsic hierarchical nature of carbon nanotube random networks owed to a dry-induced out-of-plane self-assembly phenomenon. Moreover, static contact angles of sessile water drops cast on carbon nanotube composite surfaces were measured, finding that our SWCNT random network films are more hydrophobic than our MWCNT random network films. This behavior may be ascribed to remarkable differences in the two film morphology induced by our preparation method. However, since the characteristic dimension of SWCNT is one order of magnitude smaller than MWCNT, when a SWCNT film is placed on a MWCNT film an extrinsic hierarchical morphology occurs making the resulting composite surface highly hydrophobic ($\theta=129^{\circ}\pm8^{\circ}$). We showed that our results are due to two main reasons: (i) the characteristic dimension of the self-assembly micro-structures in the SWCNT/MWCNT samples are comparable with those of micro-papillae in hydrophobic plant leaves. (ii) The hierarchical surface morphology lead to the formation of a consistent amount of air pockets, as a consequence of the transition from the hydrophobic Wenzel state to the hydrophobic Cassie-Baxter metastable state. In addition, we observed that the latter state is fairly stable in time.\\
\indent Such highly hydrophobic hierarchical carbon nanotube coatings may be very attracting for several industrial applications such as waterproof surfaces \cite{Georgakilas2008}, anti-sticking \cite{Wang2007}, anti-contamination \cite{XingJiuHuang2007}, self-cleaning \cite{Furstner2005}, anti-fouling \cite{Zhang2005}, anti-fogging \cite{Lai2012}, low-friction coatings \cite{Jung2009}, adsorption \cite{Li2010a}, lubrication \cite{Adamson1997}, dispersion \cite{Gennes2003}, and self-assembly \cite{Huang2012}. 
\section{Experimental}
\label{sec:experimental}
\subsection*{Fabrication of carbon nanotube films}
\label{sec:fabrication}
\indent Highly pure SWCNT powder (Sigma-Aldrich, assay $>$ 90\%, diameter: 0.7-0.9 nm) and MWCNT powder (Nanocyl, NC7000, assay $>$ 90\%, diameter: 5-50 nm) were dispersed in aqueous solution (80 $\mu$g\:mL$^{-1}$) with 2\% w/v sodium-dodecil-sulfate (Sigma-Aldrich, assay $>$ 98.5\%) anionic surfactant. In addition, to better disperse the suspension, carbon nanotubes were tip-ultrasonicated (Branson S250A, 200 W, 20\% power, 20 KHz) in an ice-bath for an hour and the unbundled supernatant was collected by pipette. The result was a well-dispersed suspension which is stable for several months. Carbon nanotube films were fabricated by a vacuum filtration process of 1 mL in volume of the dispersion cast on mixed cellulose ester filters (Pall GN6, 1 in diameter, 0.45 $\mu$m pore diameter). In order to prepare hierarchical MWCNT/SWCNT films, after filtering 1 mL in volume of SWCNT dispersion, 1 mL in volume of MWCNT dispersion was filtered. This process occurred also in reverse order to produce SWCNT/MWCNT films. In this way, a stack of two different film layers were obtained. Subsequently, rinsing in water and in a solution of ethanol, methanol and water (15:15:70) to remove as much surfactant as possible was performed. Samples were made uniformly depositing by the dry-transfer printing method carbon nanotube films on Carlo Erba soda-lime glass slides. More details about this novel deposition technique without chemical deposition processes have been reported elsewhere \cite{DeNicola2014}.
\subsection*{Sample characterization}
\label{sec:characterization}
\indent Scanning electron microscopy micrographs were acquired with Zeiss Leo Supra 35 field emission scanning electron microscope (FEG-SEM) and analyzed in order to measure carbon nanotube bundle diameter, network pore, and micro-structure feature (height and area) distributions. A statistical analysis of these quantities was performed and the values reported in Table \ref{tab:table1} were estimated by taking the quantity distribution mode values and standard deviations. In particular, we performed micro-structure area measurements analyzing with a threshold algorithm the film SEM micrographs at magnification $30,000\times$ and considering their irregular shape. The analysis of micro-structure height was carried out on SEM images acquired at magnification $10,000\times$ at grazing angle, i.e. by tilting the sample at an angle very close to $90^{\circ}$ with respect to the sample normal. In such a way, the height of film micro-structures can be estimated by trigonometric measurements. The film pore area defined as the area of the irregular empty regions delimited by the intersection among carbon nanotube bundles was quantified by the statistical analysis with a threshold algorithm of film SEM images at the highest magnification ($200,000\times$), where pores are clearly observable. The radius of the pore was calculated by considering the pore area as that of a circle.
\subsection*{Contact angle measurements}
\label{sec:spectroscopy}
\indent Images of sessile water drops cast on carbon nanotube films were acquired by a custom setup with a CCD camera. Static advanced contact angles were measured increasing the volume of the drop by step of 1 $\mu$L, and a plugin \cite{Stalder2010} for the open-source software ImageJ was exploited to estimate the contact angle values. This plugin exploits an algorithm based on a small-perturbation solution of the Young-Laplace equation. \cite{Adamson1997} Furthermore, the presented method is applied to a continuous image of the droplet by using cubic B-Spline interpolation of the drop contour to reach subpixel resolution. Every contact angle value reported is the average over 5 measures on images of droplets cast on 5 different points of the film (namely in the center, north, south, east, and west part). The deionized water (18.2 M$\Omega$\:cm) drop volume used to achieve the contact angles of samples was $V=10\;\mu$L. Moreover, every contact angle was measured $15$ s after drop casting to ensure that the droplet reached its equilibrium position.
\begin{acknowledgements}
The authors thank R. De Angelis, F. De Matteis, and P. Prosposito (Universit\'a di Roma Tor Vergata, Roma, Italy) for their courtesy of contact angle instrumentation. This project was financial supported by the European Office of Aerospace Research and Development (EOARD) through the Air Force Office of Scientific Research Material Command, USAF, under Grant No. FA9550-14-1-0047.
\end{acknowledgements}

\end{document}